\documentclass[10pt,twocolumn,letterpaper]{article}

\usepackage{cvpr}
\usepackage{times}
\usepackage{epsfig}
\usepackage{graphicx}
\usepackage{amsmath}
\usepackage{amssymb}
\usepackage{multirow}
\usepackage[T1]{fontenc}
\usepackage[polish, english]{babel}
\usepackage[utf8]{inputenc}

\cvprfinalcopy 

\def\cvprPaperID{****} 

\ifcvprfinal\fi
\begin{document}

\title{Superkernel Neural Architecture Search for Image Denoising}

\selectlanguage{polish}
\iftrue
\author{Marcin Możejko \\
{\tt\small marcin.mozejko@tcl.com}
\and
Tomasz Latkowski\\
{\tt\small tomasz.latkowski@tcl.com}
\and
Łukasz Treszczotko\\
{\tt\small lukasz.treszczotko@tcl.com}
\and
Michał Szafraniuk\\
{\tt\small michal.szafraniuk@tcl.com}
\and
Krzysztof Trojanowski\\
{\tt\small k.trojanowski@tcl.com}
}

\fi

\selectlanguage{english}

\maketitle
\thispagestyle{empty}

\begin{abstract}
   Recent advancements in Neural Architecture Search (NAS) resulted in finding new state-of-the-art Artificial Neural Network (ANN) solutions for tasks like image classification, object detection, or semantic segmentation without substantial human supervision. In this paper, we focus on exploring NAS for a dense prediction task that is image denoising. Due to a costly training procedure, most NAS solutions for image enhancement rely on reinforcement learning or evolutionary algorithm exploration, which usually take weeks (or even months) to train. Therefore, we introduce a new efficient implementation of various superkernel techniques that enable fast (6-8 RTX2080 GPU hours) single-shot training of models for dense predictions. We demonstrate the effectiveness of our method on the SIDD+ benchmark for image denoising \cite{sidd}.
\end{abstract}


\section{Introduction}\label{SEC_intro}

Neural architecture search (NAS) seeks to automate the process of choosing an optimal neural network architecture. Thanks to the trainable formulation of a model structure selection, it enables optimization not only for the underlying task but also for additional execution properties, \eg memory constraints, or inference time. A number of search techniques have been proposed, including ones based on reinforcement learning \cite{NAS_reinforcement}, evolutionary algorithms \cite{NAS_evolution} or gradient descent \cite{DARTS, fbnet, liu2019autodeeplab, cai2018proxylessnas}.

The greatest challenge posed by most NAS algorithms is a trade-off between search time and memory requirements of the search procedure. In classical approaches, namely reinforcement learning or evolutionary algorithms, the statistics collected during parallel training of multiple samples from the architecture space drive the exploration process. As the exploration algorithm usually needs to train these samples from scratch, the search procedure requires hundreds \cite{liu2017progressive} or thousands \cite{zoph2017learning, real2018regularized, chen2018searching} of GPU hours to be performed. In order to alleviate this problem, multiple techniques were proposed. One of the most popular ones either narrows down the search to find a local cell structure which is later replicated throughout the network \cite{zoph2017learning} or sharing weights across trained model samples \cite{pham_2018}. It is crucial to notice that the memory requirements of a given training procedure are proportional to the most memory consuming model from the explored search space. This property makes these methods a good first shot at performing neural architecture search for image enhancement tasks, where the size of individual models often fills the entire GPU memory.

An alternative approach, which is usually called \textit{single-shot} \cite{ONE_SHOT_2018, ONE_SHOT_SMASH, guo2019single} search, seeks a different solution to the aforementioned issues. It models the search space in the form of a \textit{supernetwork}, which is also a neural network itself. Then the search process is performed using gradient descent. These methods frequently reduce the overall search time from a couple of weeks to even a few hours \cite{single_path_nas}. Unfortunately, a huge downside of this approach is the fact that since \textit{supernetwork} contains the whole search space,  its memory requirements are significantly larger than those of typical network from the search space. This issue is even more important for the case of dense prediction tasks, where even a single model might be highly demanding in terms of memory use.

An efficient midway point is the \textit{superkernel} approach. It reduces the structural \textit{supernetwork} part to a single convolutional kernel, which cuts its memory requirements by order of magnitude. This improvement is achieved at the cost of narrowing down the search space selection to kernel sizes and the number of filters of a standard convolution operator.

A common solution to both prolonged training time and memory usage is using a \textit{proxy} dataset. This dataset is usually smaller, both in terms of the number of examples and image size (\eg using CIFAR10 instead of ImageNet for image classification). After initial training on a smaller dataset, the search procedure transfers the final architecture to the target task. Several approaches try to perform an architecture search procedure in a \textit{proxyless} manner, directly on target data. This is achieved by either using a simplified search space \cite{fbnet} or by compressing / pruning \textit{supernet} during training \cite{cai2018proxylessnas}. The proxy-less approach may be particularly appealing for image enhancement tasks (like image denoising). For these tasks, using a \textit{proxy} dataset is hard as proper training procedure requires high-resolution images.

In this work, we propose a new relaxed \textit{superkernel} solution for image denoising, which is fast in training (6-8 GPU hours), memory-efficient (a \textit{supernetwork} fits a single GPU with a batch size of 4) and might be trained in a \textit{proxyless} manner using input image resolution of 128x128. We evaluate our models on a SIDD+ \cite{sidd}, (part of NTIRE 2020 Challenge on Real Image Denoising \cite{Abdelhamed_2019_CVPR_Workshops, sidd, SIDD_2018_CVPR}) dataset for image denoising achieving state-of-the-art results.



\section{Related work}\label{SEC_related}



\subsection{Reinforcement Learning and Evolutionary algorithms for Neural architecture search}


Initial approaches tackled Neural Architecture Search by using reinforcement learning and evolutionary algorithms. A seminal paper \cite{NAS_reinforcement} used a reinforcement learning controller in order to produce network structures which were then trained from scratch and evaluated. Because of that, the whole training procedure took thousands of GPU hours to train a state-of-the-art model for the CIFAR10 dataset. In order to speed up computations, in the following work \cite{zoph2017learning}, the search space was narrowed down to only a small number of subnetwork structures, later repeated to form the final architecture. This approach was later improved by sharing weights across different subnetworks in order to transfer knowledge \cite{pham_2018}. Similarly, \cite{real2018regularized} used a genetic algorithm instead of reinforcement learning to drive the optimization process, whereas \cite{chen2018searching} combined both of these approaches.

As memory requirements of these methods are proportional to the memory requirements of the biggest model in search space, this technique was used for memory demanding image enhancement tasks like super-resolution \cite{guo2020hierarchical, song2019efficient, superresolution-chu}, medical image denoising \cite{LIU2019306}, image restoration \cite{zhang2019memoryefficient, restorationgerard, ho2020neural} and image inpainting \cite{Li2019ArchitectureSF}.

\subsection{\textbf{\textit{Single-shot}} approaches}

Another approach is to model the search space itself in the form of a neural network. Usually, it is done by a continuous sampling relaxation procedure, which approximates the discrete process of architecture selection in a differentiable manner. \eg in \cite{DARTS} authors used softmax weights to approximate operation selection. More precisely, the approximation was achieved by combining outputs from each operation using softmax weights with learnable logits. The \textit{supernetwork} was trained using a two-level gradient-based optimization process to separate training of convolutional weights and structural parameters. The final architecture was obtained by choosing operations by the highest logit value.

Authors of \cite{liu2019autodeeplab} pushed this approach even further, and used softmax weights as a relaxation of both global CNN architecture (channels, strides, depth, connectivity) and local cell structure. The algorithm finds a network for a semantic image segmentation task that adjusts its architecture to a provided dataset. A search space forms a grid built of a local layer structure called a cell. Each cell is a general acyclic graph with learnable connections between $k$ individually chosen and optimized operations. A grid consists of a replicated cell and a connectivity structure, which enables modeling multiple popular CNN designs like DeepLabv3, AutoEncoder, or Stacked Hourglass. The architecture search phase lasts several days on a P100 GPU.

Given that softmax weights can be an inaccurate approximation of discrete sampling when the entropy of modeled distribution is high, \cite{fbnet} introduced Gumbel Softmax approximation to neural architecture search \cite{jang2016categorical}. In this technique, stochastic weights, used for combining different operations, seem to resemble one-hot coefficients better than softmax. The proposed solution is a computationally light-weight algorithm that finds a device-aware CNN architecture for an image classification task. The algorithm works on a predefined global CNN structure, /ie, channels, strides, and depth. The search space is spanned by a set of tensor operations, including different convolution setups, pooling, and skip connections. Optimal operation is found via gradient descent for each layer individually, taking into account classification metrics as well as \texttt{FLOP}s. The algorithm finds different architectures for different hardware setups, i.e., Samsung S10, and iPhone X. Again, the architecture search phase is time-consuming at hundreds of GPU hours.

The massive downside of the \textit{single-shot} methods presented above is that they require computations of all possible operations in order to perform a single iteration of the search. In \cite{cai2018proxylessnas}, authors try to circumvent this issue by introducing search space subsampling, which decreases memory cost. At each iteration, a training update is performed only on a randomly selected subset of possible operations. Still, these models usually require approximately twice as much memory as the most demanding model from the search space.

\subsection{Superkernels}

In order to alleviate memory issues, \cite{single_path_nas} introduces a new technique called a \textit{superkernel}. In this method, the authors switch the search procedure from operation selection to kernel selection. It uses a concrete distribution in order to zero out parts of the maximal kernel, which is trained and shared across all models from the search space. As the size of the output of the convolution operation is usually of a few orders of magnitude greater than the size of its parameters, the additional cost of a search performed in the kernel space is negligible. In \ref{SEC_nas}, we introduce a few new variations of this technique based on the Gumbel Softmax relaxation \cite{jang2016categorical}.


\subsection{Deep Learning for Image denoising}

Several deep learning approaches were applied to image denoising task since the first successful MLP approach \cite{Schmidt_2014_CVPR} achieved performance comparable to then state-of-the-art BM3D \cite{bm3d} approach.  In \cite{zhang2017beyond}, the authors used stacked convolutional blocks in order to approximate the residuum between noisy image and its cleaned version. Similarly, \cite{mao2016image} introduced a residual connection not only between the input and the output of a network but also between model chained convolutional blocks. In Memnet \cite{memnet}, the authors introduced memory mechanisms based on \textit{recurrent} and \textit{gating} units in order to combine multi-level representations of an input image.

All of the methods presented above do not use any form of pooling/downsampling that introduces a high computational cost. In order to alleviate this issue \cite{liu2018multi} used U-NET architecture. By applying the pooling / upsampling operations, they decreased the resolution of the inner convolutional volumes. It significantly cut the memory and computational burden of computations. In SGN \cite{Gu_2019_ICCV}, the authors noticed that using traditional pooling / upsample operations significantly decreased the quality of reconstruction as a lot of low-level image information is lost. Because of that, they introduced a shuffling (subpixel) rescaling designed to keep more details without significant growth in a computational cost. Moreover - thanks to a \textit{top-down} self-guidance algorithm enables a light-weight combination of the multiscale image features.


\section{Models for image denoising}\label{SEC_models}

In this section, we will introduce the base architectures that we used in our experiments. We transformed each of them into Neural Architecture Search \textit{supernetworks} by introducing \textit{superkernels} instead of standard convolutional filters.

\subsection{Superkernel-based Multi Attentional Residual U-Net}\label{tl-superkernel-model}

\begin{figure}[t]
\caption{The architecture of \textbf{Superkernel-based Multi Attentional Residual U-Net} network}
\centering
\includegraphics[width=0.45\textwidth]{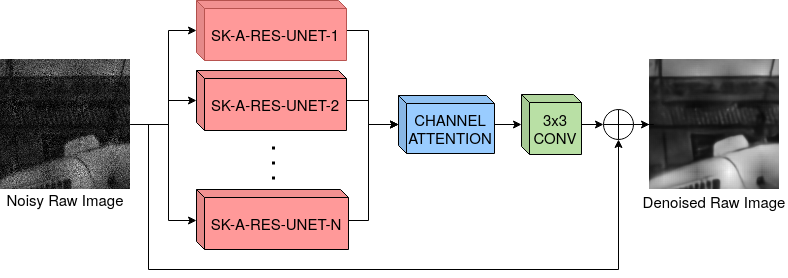}
\label{fig:skmarunet}
\end{figure}

The proposed architecture of the \textbf{Superkernel-based Multi Attentional Residual U-Net} network is shown in Figure \ref{fig:skmarunet}. The network comprises multiple subnetworks named \texttt{SK-A-RES-UNET} trained simultaneously. The outcome of each subnetwork is passed to the channel attention block \cite{2018ChannelAttention}. As the final layer, we used convolution with a kernel size of $k=3$ to reduce the number of channels. The final output is additionally summed up with the input image.

\begin{figure}[b]
\caption{The architecture of \textbf{Superkernel-based Attentional Residual U-Net} subnetwork}
\centering
\includegraphics[width=0.45\textwidth]{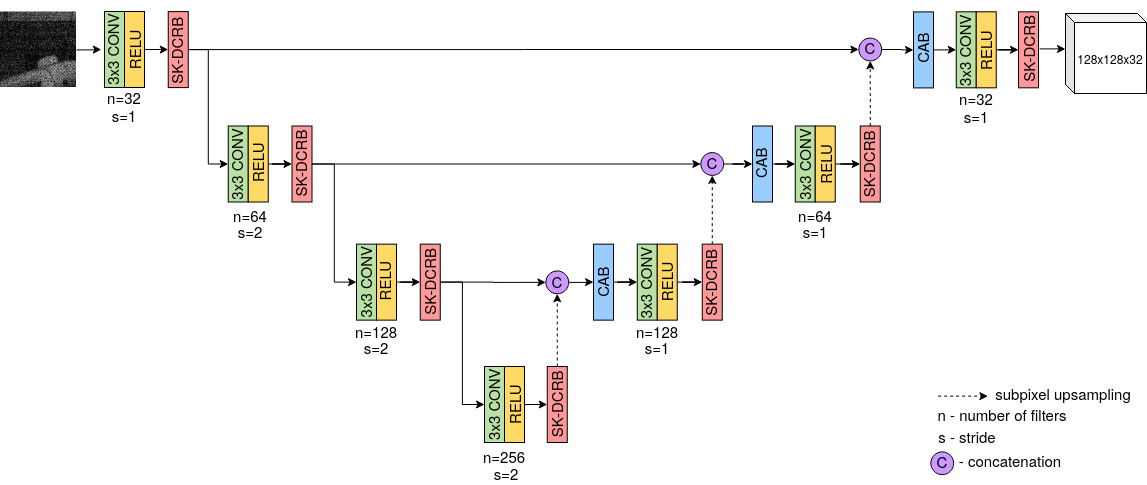}
\label{fig:skarunet}
\end{figure}

The architecture of \textbf{Superkernel-based Attentional Residual U-Net} subnetwork is shown in Figure \ref{fig:skarunet}. This network is a modification of U-Net architecture, additionally equipped with attention mechanism and superkernel-based densely connected residual blocks (SK-DCRB). Each encoder layer has the same structure. Specifically, it consists of a convolutional layer with kernel size $k=3$ followed by the ReLU activation function and SK-DCR block. As in vanilla U-Net architecture, the output of the encoder layer is passed to the decoder layer at the same spatial level. Each decoder layer takes the output from skip connection and the result from the previous layer. Unlike the common U-Net architectures, after the concatenation operation in each decoder layer, the channel attention block (CAB) is applied. We used the convolutions with a stride of 2 for down-sampling and shuffle / subpixel layers \cite{Gu_2019_ICCV} for up-sampling.

\begin{figure}[t]
\caption{The outline of \textbf{Superkernel Densely Connected Residual Block }}
\centering
\includegraphics[width=0.45\textwidth]{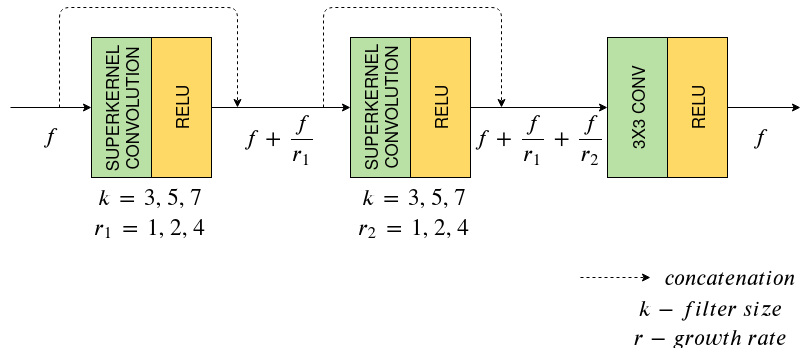}
\label{fig:skdcrb}
\end{figure}

Figure \ref{fig:skdcrb} presents the schema of \textbf{Superkernel Densely Connected Residual Block}. This block has a similar structure as proposed in \cite{DCHN}. Each SK-DCR block comprises three convolutional layers followed by ReLU activation functions. We substituted the first two convolutional operations with superkernel. This approach allows the network to learn the most appropriate kernel size and growth rate. The output convolutional layer restores the number of filters.


\subsection{Superkernel SkipInit Residual U-net}\label{lt-superkernel-model}

\begin{figure}[t]
\caption{The architecture of \textbf{Densely Connected Residual Block with SkipInit Connection} subnetwork}
\centering
\includegraphics[width=0.45\textwidth]{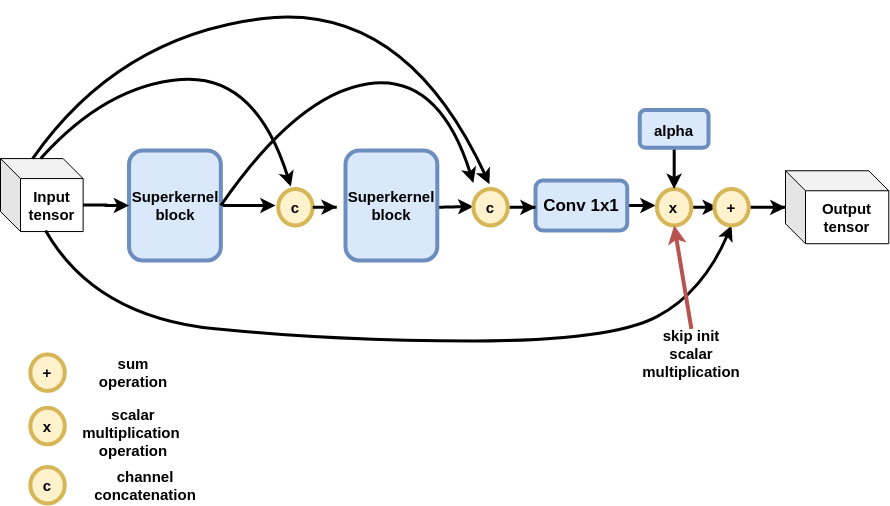}
\label{fig:skip-init-dcr}
\end{figure}

The architecture of the network is similar to the one delineated in Figure \ref{fig:skarunet}. The part that differs is the actual convolutional blocks. In \textbf{Superkernel-based SkipInit Residual U-net} there are multiple (2 in the chosen network) \textbf{Densely Connected Residual Blocks}  (see for instance \cite{zhang2018residual}) with a \textbf{SkipInit} scalar multiplication to stabilize learning as proposed in \cite{de2020batch}. The outline of this block is presented in Figure \ref{fig:skip-init-dcr}. Briefly, the output of each residual branch is multiplied by a trainable scalar, which is initialized to zero, thus making the block working like identity transformation at the beginning. If the optimization chooses to use the residual branch, then this scalar (denoted by \textbf{alfa} in Figure \ref{fig:skip-init-dcr}) will alter. The number of filters at a particular level is double the number of filters at the level directly above it, starting with 64 at level $0$. The training of convolutional hyperparameters follows the Joint Superkernel regime described in Section \ref{pr:joint}.

\section{Superkernels}\label{SEC_nas}


In each of these architectures, we applied a neural architecture search in order to find the optimal number of filters and kernel sizes for each convolution block. The search procedure was based on multiple modifications of a \textit{superkernel} technique, namely:
\begin{itemize}
    \item Factorized Gumbel Superkernel,
    \item Joint Gumbel Superkernel,
    \item Filterwise Gumbel Superkernel,
    \item Filterwise Attention-based Superkernel.
\end{itemize}
In each of these methods, the neural architecture search explores different possibilities of kernel sizes and the number of filters. Each selection is obtained by appropriate slicing of a maximal kernel, called \textit{superkernel}, and the way how this slicing is performed differs across different methods. In each, the algorithm aims to optimize the structural distribution over a set of possible slices. We will later refer to the procedure of optimization of these structural parameters as \textit{search}. Below one may find a detailed description of each NAS method.

\subsection{Joint Superkernel}\label{pr:joint}

\begin{figure}[t]
\caption{\textbf{The schema of Joint Superkernel and Factorized Superkernel slice generation.} A slice generated from choosing kernel size 1 from possible $k\in\{1, 3\}$ and filter size 32 from possible $f\in\{32, 48\}$.}
\centering
\includegraphics[width=0.4\textwidth]{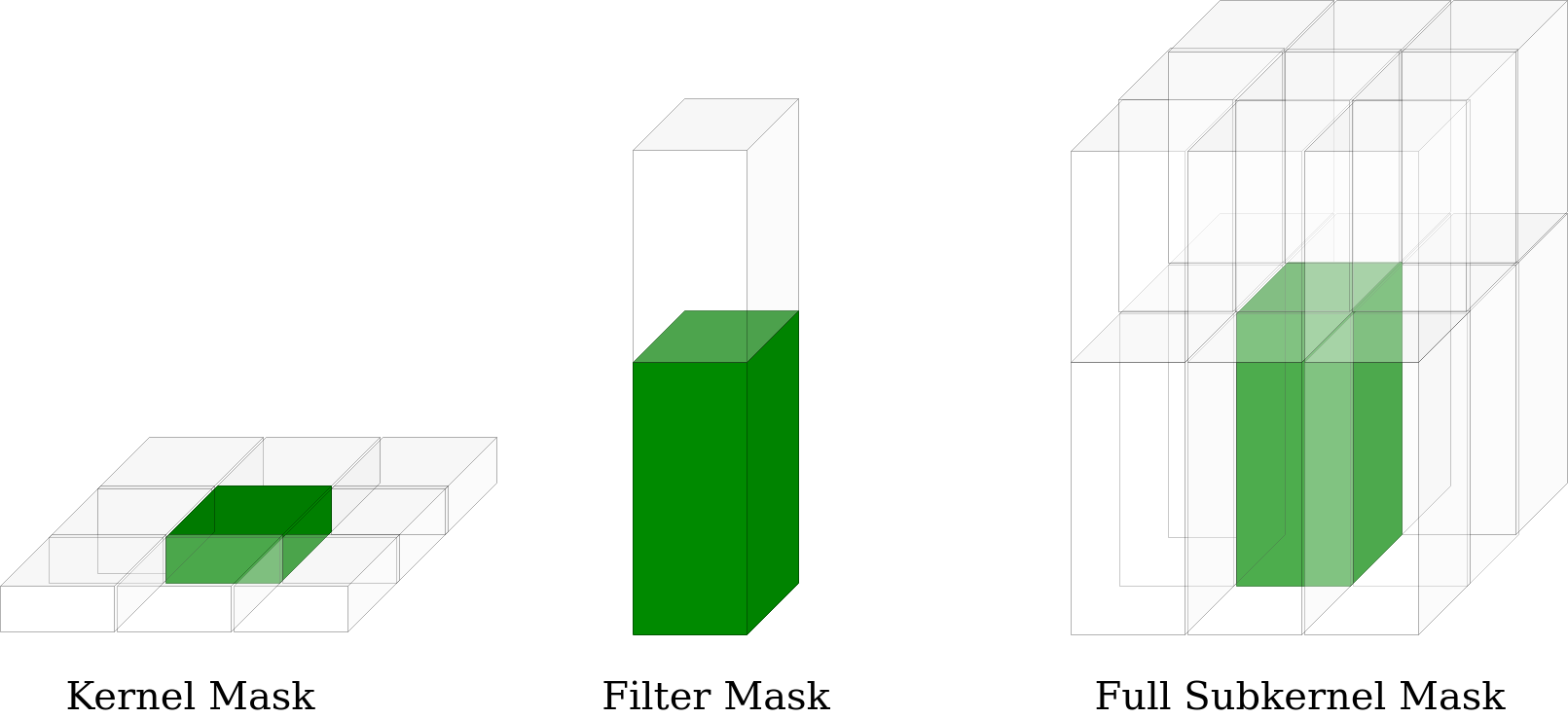}
\label{fig:sk}
\end{figure}

Using this technique, search is performed over a set of slices $S_{k, f}$ where $k\in\{k_1, \dots k_{n_k}\}$ is a size of a kernel and $f\in\{f_1, \dots, f_{n_f}\}$  is a number of filters. Both values $f_i$ and $k_i$ differ across different models and architectures. For each of these superkernels a maximal \textit{superkernel} with a kernel size of $\max_{i = 1, \dots n_k} k_i$ and $\max_{i = 1, \dots n_f} f_i$ filter numbers is trained and shared across every subkernel. A slice $S_{k ,f}$ consists of a centered subkernel with a kernel size $k$ and first $f$ filters of the maximal superkernel (see Figure \ref{fig:sk}).

The structural distribution is modeled as a softmax distribution over a set of all possible tuples $\left\{(k, f): k \in \{k_1, \dots, k_{n_k}\}, f \in \{f_1, \dots, f_{n_f}\}\right\},$ with a single logit parameter $\theta_{(k, f)}$ per tuple. 

\subsection{Factorized Superkernel}\label{pr:factorized}

In this technique the set of subkernels is the same as for Joint Superkernel in Sec \ref{pr:joint}, but the structural distribution is factorized into two independent distributions $p_{k, f} = p_k p_f,$ where $p_k$ and $p_f$ are softmax distributions over a set of possible kernels and filter sizes. In this scenario the optimized parameters form two sets of logit parameters: $\left\{\theta_{k_i}: k_i \in \{ 1, \dots n_k\}\right\}$ for kernel slices and $\left\{\theta_{f_i}: f_i \in \{ 1, \dots n_f\}\right\}$ for filter slices. In comparison to Joint Superkernel, such factorization significantly reduces number of parameters of structural distributions (from $n_f n_k$ to $n_f + n_k$) at the cost of smaller modeling flexibility. 

\subsection{Filterwise Superkernel}\label{pr:filterwise}

\begin{figure}[t]
\caption{\textbf{The schema of Filterwise Superkernel and Filterwise Attention-based Superkernel slice generation.} A slice generated by choosing kernel size 1 from possible $k\in\{1, 3\}$ and with selected filters $f\in \{2, 3, 7, 9\}$ from 11 possible.}
\centering
\includegraphics[width=0.4\textwidth]{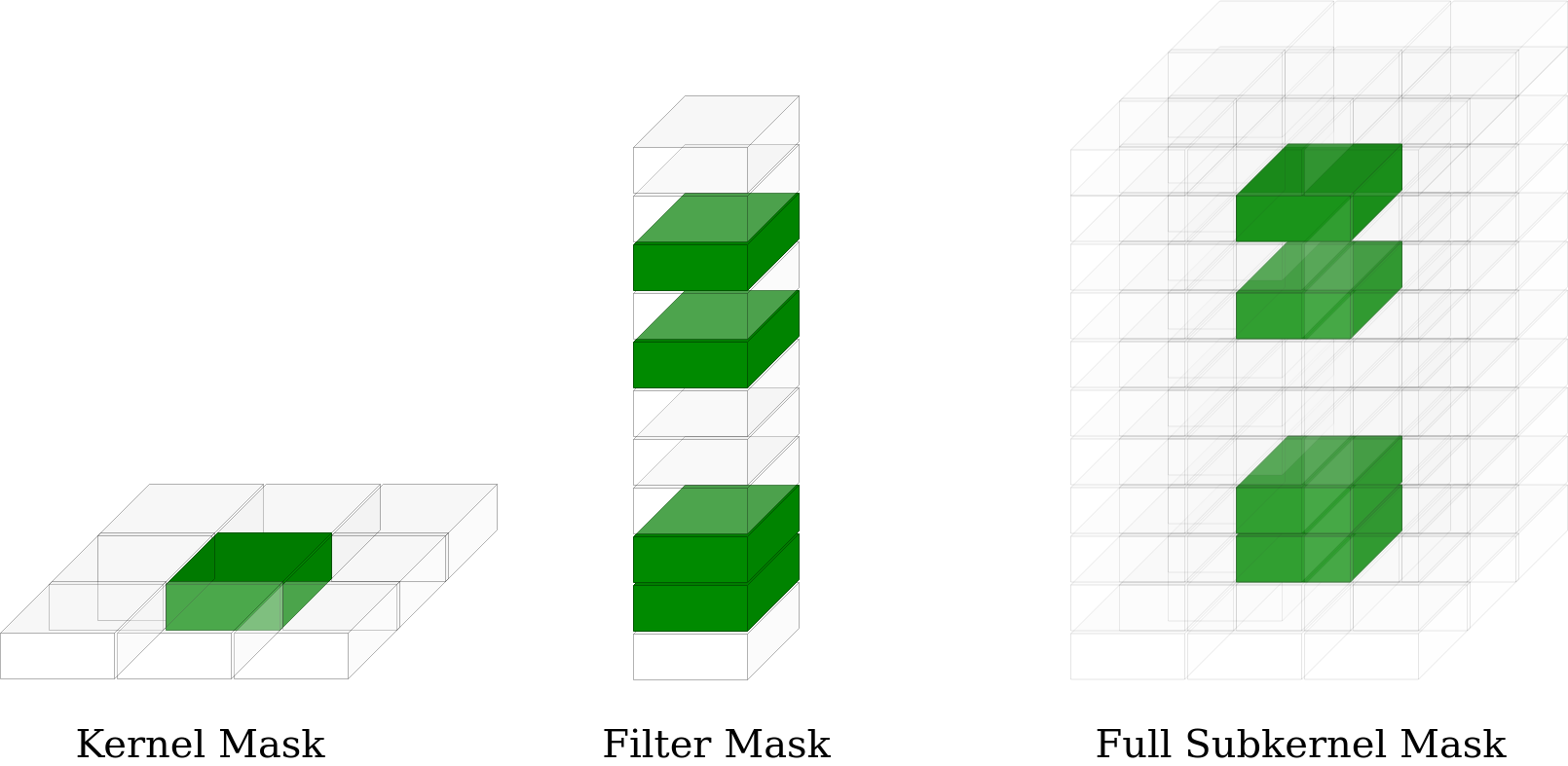}
\label{fig:fsk}
\end{figure}

In this technique the search is performed over a set of slices $S_{k, M}$ where $k\in\{k_1, \dots, k_n\}$ - is a size of a kernel and $M \in \{0, 1\}^F$ is the set of all possible subsets of $F$ potential filters. For each subkernel a maximal \textit{superkernel} with kernel size of $\max k_i$ and $F$ filters is trained and shared across every subkernel. A slice $S_{k, M}$ consists of a centered subkernel with kernel size $k$ and and those filters for which $M_i = 1$ (see Figure \ref{fig:fsk}).

The structural distribution is factorized into two independent distributions $p_{k, M}=p_k p_M$ where $p_k$ is a softmax distribution over the set of possible kernels and $p_M$ is modelled by $F$ $i.i.d.$ Bernoulli distributions. $p_k$ distribution is parametrized by a set of logit parameters $\left\{\theta_{k_i}: i \in \{1, \dots, k_n\}\right\}$, where each of $F$ independent Bernoulli filterwise distribution over $M_i$ is controlled by a single logit parameter $\theta_{f_i}.$ .

\subsection{Filterwise Attention-based Superkernel}\label{pr:fask}

In this technique, the set of subkernels is the same as in Filterwise Superkernel (Sec. \ref{pr:filterwise}). Similarly, the distribution over kernel size $k$ is also a softmax distribution over possible kernels set. A substantial difference is in the way the distributions over masks $M\in\left\{0, 1\right\}^F$ are governed. Once again, each distribution over mask $M_i$ is independent Bernoulli. This time, however, each of these distributions is parametrized by a single base logit parameter $\theta^b_{f_i}$ and an attention key vector $v^{A}_i \in \mathbb{R}^l$. The final logit of Bernoulli distribution $\theta_{f_i}$ is computed according to attention mechanism \cite{attention}:
\begin{equation}
    \theta_{f_i} =softmax\left(v^{A}(v^{A})^T\right)\theta^{b}_{f_i}.
\end{equation}

\subsection{Details of superkernel implementation}



\subsubsection{Sampling Relaxation}\label{sampling_relaxation}


As slicing is a non-continuous operation, structural parameters of networks cannot be directly optimized using gradient descent. In order to enable this type of training, we applied a continuous approximation of sampling, namely Gumbel Softmax and Relaxed Bernoulli distributions \cite{jang2016categorical}. We present an outline of this relaxation for the case of Factorized Superkernel.

We can rewrite a convolution on a volume $x$ with a sampled subkernel $S_{k, f}$ in the following manner:
\begin{equation}
    Conv\left(S_{k, f}, x\right) = \sum_{S_{k', f'}} Conv\left(S_{k', f'} , x\right)\mathbb{I}_{(k, f) = (k', f')},    
\end{equation}
where $S_{k', f'}$ goes through the set of all possible superkernels and $\mathbb{I}$ is an indicator function. The Gumbel relaxation is equivalent to the following approximation:
\begin{multline}
    \sum_{S_{k', f'}} Conv\left(S_{k', f'} , x\right)\mathbb{I}_{(k, f) = (k', f')} \approx \\
    \approx \sum_{S_{k', f'}} Conv\left(S_{k', f'} , x\right)GS(k', f'),
\end{multline}
where $GS$ is a sample from a Gumbel Softmax distribution with $\theta_{(k, f)}$ logit parametrization of a softmax distribution over a set of possible slices tuples (see Section \ref{pr:joint}).


\subsubsection{Mask sampling reparametrization}


In the approximation presented above, one still needs to compute $n_k n_f$ operations in order to perform the final summation. We decided to decrease the computational and memory burden by using a mask sampling trick. A superkernel slice might be reparametrized as $S_{k, f} = S * I_{k, f}$ where $S$ is a superkernel shared across every subkernel, $I_{k, f}$ is the appropriate subkernel slice applied to \textit{all-ones} kernel of the same shape as $S$ and $*$ is the Hadamard product of two tensors. This reparametrization enables the following reformulation of the sampling operation:
\begin{multline}
\sum_{S_{k', f'}} Conv\left(S_{k', f'} , x\right)GS(k', f') = \\
= Conv\left(S*\sum_{S_{k', f'}} I_{k', f'} GS(k', f'),  x\right).
\end{multline}
The RHS of the above equation needs only a single run of convolution operation with the maximal superkernel $S$ masked with an average subkernel mask. This substantially decreases computational burden, both in terms of \texttt{FLOP}s and memory requirement, and enables full NAS treatment for image denoising.

\subsubsection{Mask Sampling reparametrization - issue with a non-linear activation function}


One may notice that the derivation above is not accurate for a search procedure which involves not only a $Conv$ operator but also a non-linear activation function. This is because of the relation:
\begin{multline}
\sum_{S_{k', f'}} f\left(Conv\left(S_{k', f'} , x\right)GS(k', f')\right) \\
 = f\left(Conv\left(S*\sum_{S_{k', f'}} I_{k', f'} GS(k', f'),  x\right)\right)
\end{multline}
holds only for additive functions $f$. Both ReLU and PReLU, which were used in our experiments, do not have this property. Because of that, we tested two possibilities:
\begin{itemize}
    \item \textbf{full} - where in order to keep computational feasibility we decided to ignore this fact and treat the RHS of the equation above as an approximation of the LHS,
    \item \textbf{separate} - where all components of the LHS were computed separately. 
\end{itemize} 
We have tested both approaches on a set of smaller architectures and did not notice any significant difference. If a model has a separate component in its name, it indicates it was trained using a separate approach. Otherwise it was trained using full treatment.

\begin {table*}[t]
\caption {Superkernel-based SkipInit Residual U-Net model with two different architectures (with self-ensemble).} 
\label{tab:lt-self-ensemble} 
\begin{center}
\begin{tabular}{|c|c|c|c|c|}
    \hline
    \multirow{3}{*}{\textbf{Superkernel Type}} &
    \multicolumn{4}{c|}{Network architecture} \\ \cline{2-5}
    & \multicolumn{2}{c|}{Model 1} & \multicolumn{2}{c|}{Model 2} \\ \cline{2-5} 
    & \textbf{PSNR} & \textbf{SSIM} & \textbf{PSNR} & \textbf{SSIM} \\
    \hline \hline
    no superkernel & $52.6738$ & $0.99170$ & $52.6883$ & $0.99170$ \\ 
    \hline
    full & $52.7330$ & $0.99183$ & $\mathbf{52.7344}$ & $\textbf{0.99183}$ \\ 
    \hline
    separate & $\mathbf{52.74498}$ & $\mathbf{0.99184}$ & & \\ 
    \hline
    filterwise & $52.6444$ & $0.99179$ & $52.6817$ & $\textbf{0.99183}$ \\
    \hline
    filterwise with attention & $52.7059$ & $0.99173$ & $52.7053$ & $0.99191$\\ 
    \hline
\end{tabular}
\end{center}
\end{table*}

\begin {table*}[t]
\caption {Superkernel-based SkipInit Residual U-Net model with two different architectures (without self-ensemble).} 
\label{tab:lt-no-self-ensemble} 
\begin{center}
\begin{tabular}{|c|c|c|c|c|}
    \hline
    \multirow{3}{*}{\textbf{Superkernel Type}} &
    \multicolumn{4}{c|}{Network architecture} \\ \cline{2-5}
    & \multicolumn{2}{c|}{Model 1} & \multicolumn{2}{c|}{Model 2} \\ \cline{2-5} 
    & \textbf{PSNR} & \textbf{SSIM} & \textbf{PSNR} & \textbf{SSIM} \\
    \hline \hline
    no superkernel &  $52.4937$ & $\textbf{0.99157}$ & $52.4490$ & $0.99140$  \\ 
    \hline
    full & $52.4638$ & $0.99140$ & $\textbf{52.4752}$ & $0.99144$ \\ 
    \hline
    separate & $\textbf{52.4957}$ & $0.99148$ & & \\ 
    \hline
    filterwise & $52.4561$ & $0.99147$ &  $52.4151$ & $\textbf{0.99145}$ \ \\ 
    \hline
    filterwise with attention & $52.4239$ & $99138$ & $52.4278$ & $0.99152$\\ 
    \hline
\end{tabular}
\end{center}
\end{table*}


\subsubsection{Bias sampling}


We have applied a similar sampling technique as in the case of convolutional kernels to the biases. As the biases vector size is equal to the number of filters, the appropriate mask for bias slice is obtained by averaging kernel mask in all but filter dimensions.


\subsubsection{Final model distillation}


Once structural parameters and weights of a model are trained, we need to distill the model with the best kernel and filter sizes. We applied the following distillation strategy:
\begin{itemize}
    \item for structural parameters governed by a \textbf{softmax} distribution, an option with a maximal probability was chosen, 
    \item for structural masks governed by a \textbf{Bernoulli} distribution, a filter was selected if its logit was greater than $0.5$.
\end{itemize}
This distillation procedure was done independently for each superkernel. We plan to use more sophisticated distillation strategies as part of our future work.


\section{Experimental results}\label{SEC_results}


We performed several experiments in raw RGB image denoising task in order to test whether our approach to the search for the best kernel size/number of output channels combination provides noticeable benefits to the baseline (no-NAS) solution developed for the same task. In this section, we briefly describe the training procedure and the models used. We also provide the final results in terms of achieved PSNR and SSIM scores.

\begin {table*}[t]
\caption {Superkernel-based Multi Attentional Residual U-Net model for different number of subnetworks and superkernel types (without self-ensemble).} 
\label{tab:tl-1} 
\begin{center}
\begin{tabular}{|c|c|c|c|c|c|c|} 
    \hline
    \multirow{3}{*}{\textbf{Superkernel Type}} &
    \multicolumn{6}{c|}{Number of subnetworks $n$} \\
    \cline{2-7}
    & \multicolumn{2}{c|}{$n$=1} & \multicolumn{2}{c|}{$n$=3} & \multicolumn{2}{c|}{$n$=6} \\ \cline{2-7}
    & \textbf{PSNR} & \textbf{SSIM} & \textbf{PSNR} & \textbf{SSIM} & \textbf{PSNR} & \textbf{SSIM} \\
    \hline \hline
    no superkernel & $52.3517$ & $\mathbf{0.99156}$ & $\mathbf{52.4523}$ & $0.99126$ & $\mathbf{52.4833}$ & $0.99136$\\ 
    \hline
    factorized & $52.2690$ & $0.99124$ & $52.3799$ & $0.99134$ & $52.4032$ & $\mathbf{0.99150}$\\ 
    \hline
    joint & $\mathbf{52.3925}$ & $0.99138$ & $52.4153$ & $0.99140$ & $52.4118$ & $0.99147$\\ 
    \hline
    filterwise & $52.2887$ & $0.99121$ & $52.3503$ & $\mathbf{0.99145}$ & $52.3516$ & $0.99131$\\ 
    \hline
    filterwise with attention & $52.2932$ & $0.99122$ & $52.4365$ & $0.99144$ & $52.4195$ & $0.99119$\\ 
    \hline
    \multirow{2}{*}{$\mu$ $\pm$ $\sigma$} & $52.3190$ & $0.99132$ & $52.4069$ & $\mathbf{0.99138}$ & $\mathbf{52.4139}$ & $0.99137$\\
    & $\pm0.0459$ & $\pm13.39e$-$5$ & $\pm0.0373$ & $\pm7.05e$-$5$ & $\pm0.0421$ & $\pm11.22e$-$5$\\ 
    \hline
\end{tabular}
\end{center}
\end{table*}

\begin {table*}[t]
\caption {Superkernel-based Multi Attentional Residual U-Net model for different number of subnetworks and superkernel types (with self-ensemble).} 
\label{tab:tl-2} 
\begin{center}
\begin{tabular}{|c|c|c|c|c|c|c|}
    \hline
    \multirow{3}{*}{\textbf{Superkernel Type}} &
    \multicolumn{6}{c|}{Number of subnetworks $n$} \\ \cline{2-7}
    & \multicolumn{2}{c|}{$n$=1} & \multicolumn{2}{c|}{$n$=3} & \multicolumn{2}{c|}{$n$=6} \\ \cline{2-7} 
    & \textbf{PSNR} & \textbf{SSIM} & \textbf{PSNR} & \textbf{SSIM} & \textbf{PSNR} & \textbf{SSIM} \\
    \hline \hline
    no superkernel & $52.4470$ & $0.99157$ & $52.5969$ & $0.99150$ & $52.6642$ & $0.99159$\\ 
    \hline
    factorized & $52.4998$ & $0.99157$ & $52.6184$ & $0.99167$ & $\mathbf{52.6981}$ & $\mathbf{0.99176}$\\ 
    \hline
    joint & $\mathbf{52.5668}$ & $\mathbf{0.99163}$ & $52.6016$ & $0.99164$ & $52.6189$ & $0.99159$\\ 
    \hline
    filterwise & $52.5144$ & $0.99150$ & $52.5982$ & $0.99180$ & $52.6654$ & $0.99163$\\ 
    \hline
    filterwise with attention & $52.5438$ & $0.99157$ & $\mathbf{52.6880}$ & $\mathbf{0.99186}$ & $52.5943$ & $0.99150$\\ 
    \hline
    \multirow{2}{*}{$\mu$ $\pm$ $\sigma$} & $52.5144$ & $0.99157$ & $52.6206$ & $\mathbf{0.99169}$ & $\mathbf{52.6482}$ & $0.99161$\\
    & $\pm0.0409$ & $\pm4.12e$-$5$ & $\pm0.0346$ & $\pm12.64e$-$5$ & $\pm0.0369$ & $\pm8.45e$-$5$\\ 
    \hline
\end{tabular}
\end{center}
\end{table*}

\subsection{Training details}

We trained all our models on the SSID+ \cite{sidd} training data provided by NTIRE2020 Real Image Denoising rawRGB competition organizers. In our training procedure, we split the challenge images into two parts: 90\% of images for a training set, and the rest 10\% for validation. The split is stratified with respect to dataset mobile phone types.  We used the validation data provided by the organizers of the competition as the final test set. We cut the training and validation images into 128x128 patches. We use Adam optimizer \cite{adam} with a learning rate of $2*10^{-5}$ for model training. The learning rate was kept constant during the learning process. As our pipeline comprised of a large set of models of different architectures and capacities, we decided to use a popular early stopping technique to constrain the training time. During the learning process, we monitored the PSNR value on a validation dataset. We evaluated our model after every 1K steps.  In the case of no improvement in the PSNR within 30 validation evaluations, we stopped the training process. Depending on the hardware used (GeForce GTX 1080Ti or GeForce RTX 2080Ti), the training of our models lasted from a few hours up to 3 days. 


\subsection{Superkernel SkipInit Residual U-net}\label{results_lt}


The overall architecture of the models in this section is based on the U-net presented in \cite{DCHN}. Model 1 in this subsection is the one described in Section \ref{lt-superkernel-model} with U-net depth $3$, three searchable growth rates $0.2$, $0.4$ and $0.6$ in the DCR block. There are $2$ DCR blocks of depth $2$ at each U-net level (both for the encoder and decoder parts). At level $0$, the maximum number of possible output channels is $32$, and it increases with arithmetic progression at each layer. Model 2 differs from Model 1 only in the U-net depth ($2$ instead of $3$) and the maximum number of channels at level $0$ ($64$ instead of $32$). The performance of theses models can be seen in Tables \ref{tab:lt-self-ensemble} and  \ref{tab:lt-no-self-ensemble}.

\subsubsection{Chosen architecture}

The search space for both models included the choice of kernel size ($3$ or $5$) and growth rate ($0.2$, $0.4$, $0.6$) for the DCR block, see Figure \ref{fig:skip-init-dcr} and Section \ref{lt-superkernel-model}. There where $28$ DCR blocks in Model 1 and $20$ in Model 2. 

Model 1 in the \textbf{full} version of the \textbf{superkernel joint} framework, consecutively chose kernel size $5$, with only a couple of exceptions. As regards growth rates, all but three DCR blocks chose the largest growth rate possible. In the \textbf{separate} version, the story was similar, with kernel size $3$ chosen only once. 

In Model 2 with \textbf{full} version of \textbf{superkernel joint}, which had a higher baseline number of filters, but was shallower, as far as U-net architecture is concerned, chose kernel size $5$ virtually everywhere. Growth rates were a bit different, with approximately 40\% DCR blocks choosing a growth rate of $0.4$ and the rest choosing $0.6$.

\subsection{Superkernel-based Multi Attentional Residual U-Net}\label{results_tl}

Tables \ref{tab:tl-1} and \ref{tab:tl-2} contain PSNR and SSIM scores for architecture described in Section \ref{tl-superkernel-model}. We trained several models with a different number of subnetworks and superkernel types. Table \ref{tab:tl-1} lists results without self-ensemble whereas models presented in Table \ref{tab:tl-2} utilize self-ensemble technique. We use architectures with 1, 3, and 6 subnetworks. Each subnetwork has the same structure, as shown in Figure \ref{fig:skarunet}. As illustrated in Table \ref{tab:tl-2}, the application of the self-ensemble technique significantly improved the average PSNR and SSIM metrics. The best PSNR results were obtained for models containing six subnetworks. The mean PSNR value of models equipped with six subnetworks (with self-ensemble) outperformed the 1-subnetwork and 3-subnetwork models by 0.13 and 0.03, respectively.

\subsubsection{Chosen architecture}

The search space for the model consisted of kernel size and growth rate of densely connected residual blocks. For every first two convolutional layers in DCR block, the superkernel had to select a filter size in $k\in\{3, 5, 7\}$ and a growth rate in $1/r\in\{1, 0.5, 0.25\}$. The final superkernel choices differed depending on the part of the network. In the encoder part, the superkernel selected the kernel sizes of 5 and 7 and the growth rate of 1 and 0.5. In the decoder part, NAS selected the kernel sizes of 7 and the maximum number of filters ($r=1$) with only some small exceptions.


\section{Discussion}

In all architectures introduced, at least one of Neural Architecture Search methods significantly improved over no-NAS baseline. The best single model - with $52.74$ PSNR used the separate joint superkernel - and was achieved by SkipInit Residual U-net for Model 1. The difference between baseline architectures and searched ones is the most significant (approx. $0.12$ of PSNR) for small models (with single subnetwork), which is promising from the perspective of model deployment on a mobile device. 

The ablation study showed the importance of a self-ensembling technique. On average, it provided an improvement in both PSNR ($0.21$) and SSIM ($0.0002$) metrics. Interestingly, without self-ensembling, NAS models perform significantly worse as compared to their no-NAS baselines.

\section{Potential improvements and future work}

Although our search procedure shows the most significant PSNR improvement for small models, this comes at the cost of the selection of convolutional kernels with more spacious kernel sizes and higher the number of filters. However, a differentiable representation of kernel size enables the introduction of additional computational constraints to an optimization process. We expect that training equipped with additional computational regularization should result in finding models deployable on portable devices.

Our search procedure resembles variational inference methods for Bayesian Deep Learning; thus, our method might suffer from an insufficient exploration of a search space as reported in \cite{fort2019deep}. This issue might arise as our methods explore only points which are close to gradient descent trajectory, which narrows the explored region of the search space and makes it potentially sensitive to weight initialization. We are planning to explore this sensitivity as our future work. Moreover, in our training procedure, both weights and structural parameters are optimized jointly. We are planning to explore a two-level optimization (similar to \cite{DARTS}) in order to inspect how the coupling of these parameters affects the results.

As sampling within different superkernels is performed independently, there may arise issues connected to fairness \cite{fairnas}. Because of that, the network might quickly converge to a random sub-optimal solution, according to Matthew's law. We plan to test more advanced probabilistic models (\eg based on Markov Random Fields, recurrent neural networks, or attention) in order to model more advanced search spaces and perform broader exploration.

Additionally, we observed that the training procedure could have a significant impact on the final results. Specifically, models equipped with superkernel might need longer training time (higher patience) in comparison to non-superkernel models. We suspect that this might be a reason behind performance drop for NAS models without self-ensembling. We plan to investigate this topic in future work.

\section{Conclusion}

In this work, we introduced a fast and light-weight algorithm for neural architecture search for image denoising. We proved that it could achieve state-of-the-art results exceeding no-NAS solution by a significant margin. We have shown that proposed superkernel techniques can achieve results comparable to the state-of-the-art architectures for image denoising within few GPU hours.

{
\bibliographystyle{ieee_fullname}
\bibliography{publications}

}

\end{document}